\documentclass{IEEEtran}
\usepackage[utf8]{inputenc}

\usepackage{amsfonts}
\usepackage{amsmath,commath}
\usepackage{amssymb}
\usepackage{booktabs}
\usepackage{cite}
\usepackage{color}
\usepackage{colortbl}
\usepackage{enumitem}
\usepackage{epsfig}
\usepackage{epstopdf}
\usepackage{fancyhdr}
\usepackage{graphicx}
\usepackage{hyphenat}

\usepackage{ifthen}
\usepackage{mathtools}
\usepackage{multirow}
\usepackage{pdfpages}
\usepackage{rotating}
\usepackage{setspace}
\usepackage{subfigure}
\usepackage{textcomp}

\usepackage{xfrac}
\usepackage{gensymb}
\usepackage{epigraph}
\usepackage{accents}
\usepackage{CJKutf8}

\usepackage[hyphens]{url}
\usepackage{hyperref}
\hypersetup{breaklinks=true}

\begin{document}
	\title{ Fast and Robust Characterization of Dielectric Slabs Using Rectangular Waveguides}
	\author{\IEEEauthorblockN{Xuchen~Wang and Sergei~A.~Tretyakov}\\
		\IEEEauthorblockA{\textit{Department of Electronics and Nanoengineering, Aalto University}, Espoo, Finland\\
			xuchen.wang@aalto.fi}
	} 
	\maketitle

	\begin{abstract}
		Waveguide characterization of dielectric materials is a convenient and broadband approach for measuring dielectric constant. In conventional microwave measurements,  material samples are usually mechanically shaped to fit the waveguide opening and measured in closed waveguides. 
		This method is not practical for millimeter-wave and sub-millimeter-wave measurements where the waveguide openings become tiny, and it is rather difficult to shape the sample to exactly the same dimensions as the waveguide cross-section. 
		In this paper, we present a method that allows one to measure arbitrarily shaped dielectric slabs that extend outside waveguides.
		In this method, the measured sample is placed between two waveguide flanges, creating a discontinuity. The measurement system is characterized as an equivalent $\Pi$-circuit, and the circuit elements of the $\Pi$-circuit are extracted from the scattering parameters.
		We have found that the equivalent shunt impedance of the measured sample is only determined by the material permittivity and is rather insensitive to the sample shape, position, sizes, and other structural details of the discontinuity. This feature can be leveraged for accurate measurements of permittivity.
		The proposed method is very useful for measuring the permittivity of medium-loss and high-loss dielectrics from microwave to sub-terahertz frequencies.
	\end{abstract}
	
	\begin{IEEEkeywords}
		Permittivity measurement, millimeter-wave, sub-millimeter-wave, rectangular waveguide.
	\end{IEEEkeywords}

	\section{Introduction}
	
	Material characterization is an essential step in designing electromagnetic devices. The recent fast developments of wireless communications (5G and beyond) impose strong demands for the characterization of dielectric materials at millimeter-wave and even higher frequencies. 
	In general, the methods for measuring dielectric constant can be divided into two groups: one is based on resonating systems and the other is utilizing non-resonant transmission-lines structures \cite{krupka2006frequency}. 
	In resonance-based methods,  the sample is usually machined into a dielectric resonator with a high $Q$-factor. The real part of the permittivity can be predicted from the resonant frequency and the loss tangent is extracted from the $Q$-factor of the measurement system \cite{rajab2005dielectric,krupka1999use}. 
	Alternatively, one can position the sample into a high-$Q$ cavity and obtain the permittivity value by measuring the perturbation of the resonant frequency and the $Q$-factor of the system before and after loading the cavity  \cite{dudorov2005millimeter}. 
	Generally speaking, resonance methods provide the best accuracy in the estimation of both real and imaginary parts of permittivity for low-loss dielectrics.
	The drawback of this method is that the measured frequencies must be discrete, corresponding to the resonant frequencies of the system.
	In addition,
	the dimensions of the resonators become tiny at millimeter-wave or higher frequencies, which imposes considerable practical difficulties.

	\begin{figure}[h!]
		\centering
		\includegraphics[width=0.7\linewidth]{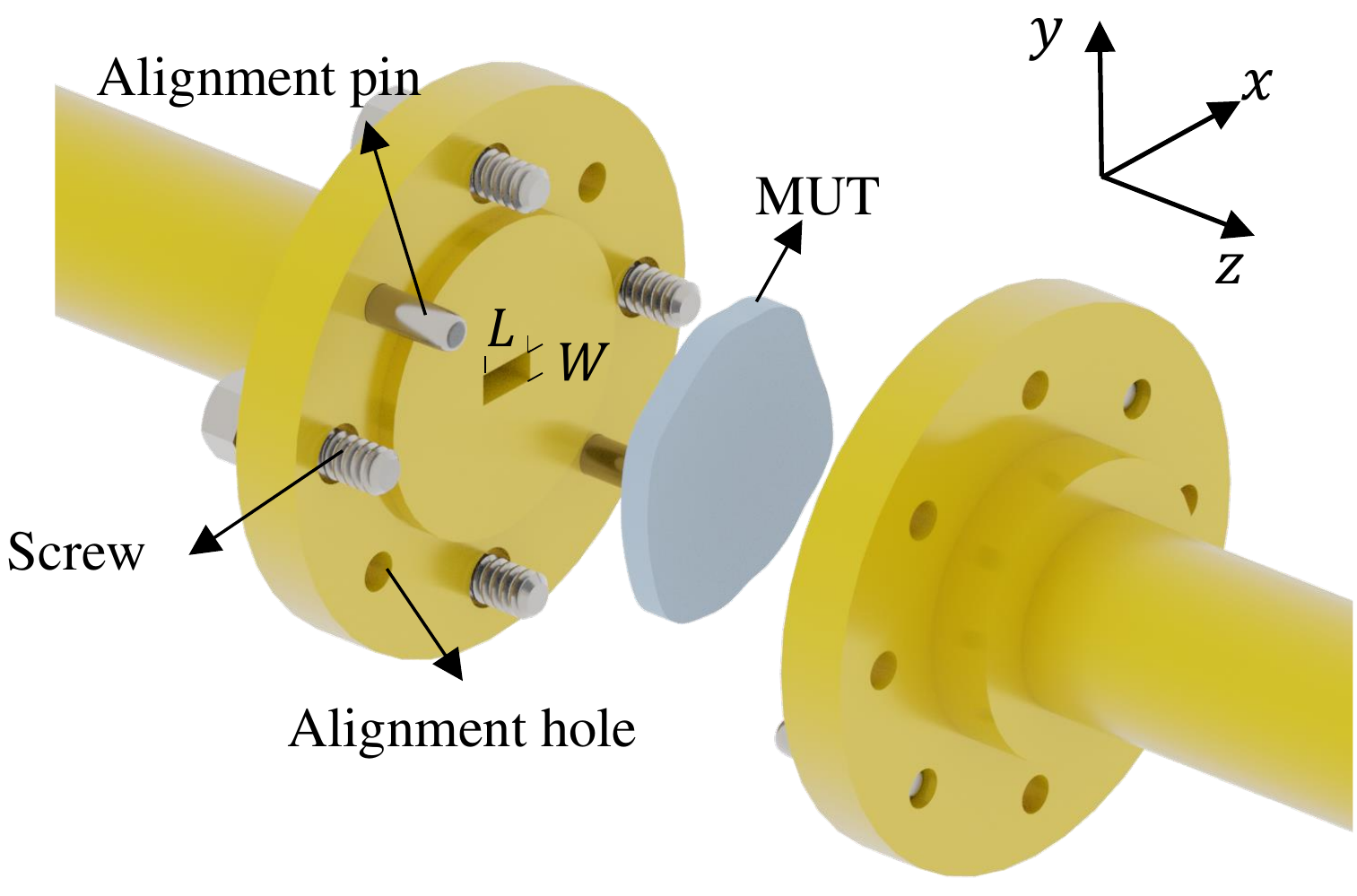}
		\caption{ Measurement setup using millimeter-wave rectangular waveguides. The sample under test (SUT) is positioned in between two waveguide flanges. The actual setup is fastened by  screws.}\label{fig:measurement_setup}
	\end{figure}
	
	In the transmission-line-based method, the measured sample is connected as a load or insertion in a waveguiding structure.
	By measuring the reflection and/or transmission coefficients ($S$-parameters) of the system, the dielectric properties of the material can be determined in a broad frequency range \cite{you2017materials}.
	Due to the absence of a setup resonance, the measured $S$-parameters are not so sensitive to the dielectric losses as in the resonator method (especially for thin dielectric samples), and therefore the measurement accuracy for the loss tangent is generally worse than in the resonator method. 
	For this reason, the transmission-line method is most suitable for the characterization of medium-loss and high-loss dielectrics \cite{krupka2006frequency,costa2017electromagnetic}. 
	Transmission-line structures can be formed by many structures, such as metallic \cite{nicolson1970measurement,weir1974automatic,hasar2009microwave} or dielectric waveguides \cite{rajab2005dielectric,nefedova2015dielectric}, coplanar waveguides \cite{deo2015microstrip}, microstrip lines \cite{janezic2003permittivity}, free space \cite{campbell1978free}, and so on. 
	One of the most commonly used methods is developed in \cite{nicolson1970measurement,weir1974automatic}, which is well known as the  Nicolson–Ross–Weir method. In this method, the sample under test (SUT) is embedded into a  rectangular waveguide and fully in contact with the waveguide walls. 
	However, in practice, there are inevitable air gaps between the surfaces of the SUT and the waveguide walls. This is a considerable restriction for measurements at millimeter-wave and above frequencies where the waveguide dimensions are of the order of millimeter or smaller \cite{sahin2020simplified}.

	In order to avoid the problem of imperfect contact with the waveguide walls, 
	it is preferable and easier to test the sample outside waveguides \cite{abbas1998rectangular, hyde2008nondestructive,hyde2015nondestructive,haro2019higher}.  
	For example, in \cite{hyde2008nondestructive,hyde2015nondestructive}, the samples are positioned between two waveguide flanges, creating a discontinuity from which the electromagnetic energy is allowed to leak away. 
	The measurement setup is then modeled in commercial simulation tools or using self-developed numerical algorithms to calculate the $S$-parameters.
	By fitting the numerically simulated $S$-parameters with the measured values, the permittivity of the sample can be estimated. 
	However, in this method, it is necessary to ensure that the actual measurement setup is accurately represented in numerical modeling, e.g., the dimensions of the waveguide flanges and the test sample, since the $S$-parameters of the system are affected by all these details. 
	Practical limitations on the modeling accuracy of all the setup details do not allow accurate millimeter-wave measurements, because the configuration of millimeter-wave waveguide flanges is usually not planar and contains other structures such as holes, chokes, alignment pins, and screws, which are difficult to model accurately (see Fig.~1). 
	Obviously, in the millimeter-wave range and above, it is more convenient to measure the sample outside waveguides, without caring about the sample shape and positioning, the flange types, and other accessories.
	
	In this paper, we propose such a method that can be used to measure arbitrary-shaped dielectric slabs outside rectangular waveguides  (see Fig.~\ref{fig:measurement_setup}). 
	We qualitatively analyze the electromagnetic fields in the discontinuity and use the understanding of field distribution to  model the discontinuity as a $\Pi$-circuit where each circuit component can be extracted from the measured $S$-parameters.  
	We have found that the equivalent shunt impedance of the discontinuity is rather insensitive to the sample shape and the structural details in the discontinuity, and it is only determined by the permittivity and thickness of the sample. 
	This feature can be readily used for the extraction of dielectric permittivity if the sample thickness is known. 
	
	The paper is organized as follows: in Section~II, we introduce the physical principle of the proposed method. In Sections~III and~IV, we separately discuss the permittivity extraction methods for electrically thin and thick dielectric slabs. The measurement uncertainties are analyzed in Section~V.

	\section{Measurement principle}
	
	In this section, we introduce the physical principle of the proposed measurement method. We start from the field analysis for the measurement setup. Then, we model the discontinuity as an equivalent circuit, and  verify stability of the equivalent shunt impedance.
	
	\subsection{Field distribution in the discontinuity}
	
\begin{figure}[h!]
		\centering
		\includegraphics[width=0.95\linewidth]{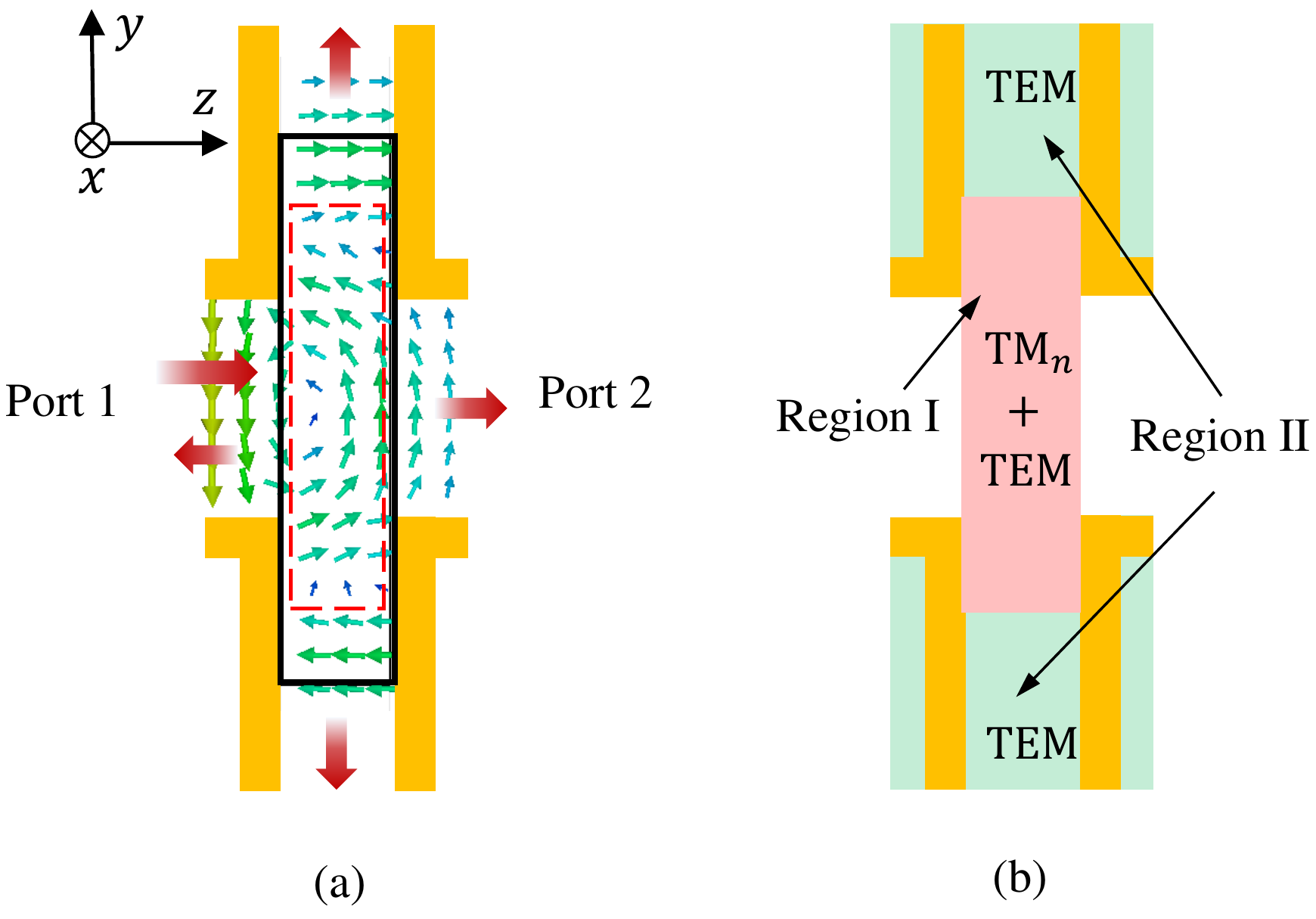}
		\caption{ (a) Vectorial field distribution in the $yz$ cross-section of the measurement setup for excitation from Port~1. In the simulations, the operating frequency is $60$~GHz, the waveguide aperture size is $L=3.76\  \rm~mm$ and $W=1.88\  \rm~mm$ (WR-15). Note that, throughout the paper, we use WR-15 waveguides for all the numerical and experimental analyses.  The dielectric slab has the permittivity of $\epsilon_{ r}=4(1-0.01j)$ and thickness of $d=625\ \mu$m. (b) Division of regions.  }\label{fig: field model}
	\end{figure}

	Figure~\ref{fig:measurement_setup} shows the actual measurement setup based on  millimeter-range rectangular waveguides. An arbitrarily shaped piece of a dielectric slab is positioned between two waveguide flanges. The waveguide aperture is fully covered by the dielectric sample. 
	Although the structure of waveguide flanges contains many small details, for the following conceptual analysis it is possible to simplify the flanges as planar metallic walls. 
	A cross-section of the discontinuity is shown in Fig.~\ref{fig: field model}(a). 
	Waves incident from Port~1 are partially reflected and absorbed by the sample (shown as a black rectangle) and partially leak to free space via the gap. The rest of the power enters Port~2. The discontinuous junction is composed of a parallel-plate waveguide (PPWG) connected with a pair of rectangular waveguides. 
	As we know, if the waveguides are continuous, and the operating frequency is below the cutoff frequencies of higher-order modes,  the rectangular waveguide only supports the TE$_{01}$ mode, and the PPWG only supports the TEM mode. However, in the junction, both rectangular waveguides and PPWG are discontinuous. To adapt to the configuration of the junction, higher-order modes of the waveguides are excited. Therefore, the fields in the junction have a complicated composition, which is a combination of waveguide fundamental modes and many higher-order modes. The fields distribution in the junction region can be rigorously computed using the  mode-matching method \cite{hyde2008nondestructive}. Figure~\ref{fig: field model}(a) shows the simulated electric field distribution in the waveguide junction. As we can see, near the rectangular waveguide apertures, strong TM$_n$ modes ($\mathbf{E}_y\neq0$) of the PPWG  are excited, and these modes continue to propagate in the PPWG along the  $y$-direction.  However, after the higher-order modes leave the junction region, they only see a continuous PPWG. Since the excitation frequency is below the cutoff frequencies of these higher-order modes in the PPWG, these modes are evanescent and decay exponentially away from the junction. At some distance (see the top/bottom edge of the red dashed rectangle), the higher-order modes become negligible and only the fundamental TEM mode continue propagation in the PPWG formed by two flanges.

	\subsection{Circuit modeling of the discontinuity}
	According to the field distribution in Fig.~\ref{fig: field model}(a), the electromagnetic environment of the discontinuity can be divided into two volumetric regions [see Fig.~\ref{fig: field model}(b)]: 
	Region~I (highlighted in pink) encloses the volume where TM$_n$ modes survive;
	Region~II (highlighted in green) includes the remaining volume and all the surroundings outside the setup. 
	Region~I ($\mathbf{E}_y\neq0$) is directly connected with the waveguide ports and therefore it can be viewed as a two-port system. 
	
\begin{figure}[h!]
		\centering
		\includegraphics[width=0.8\linewidth]{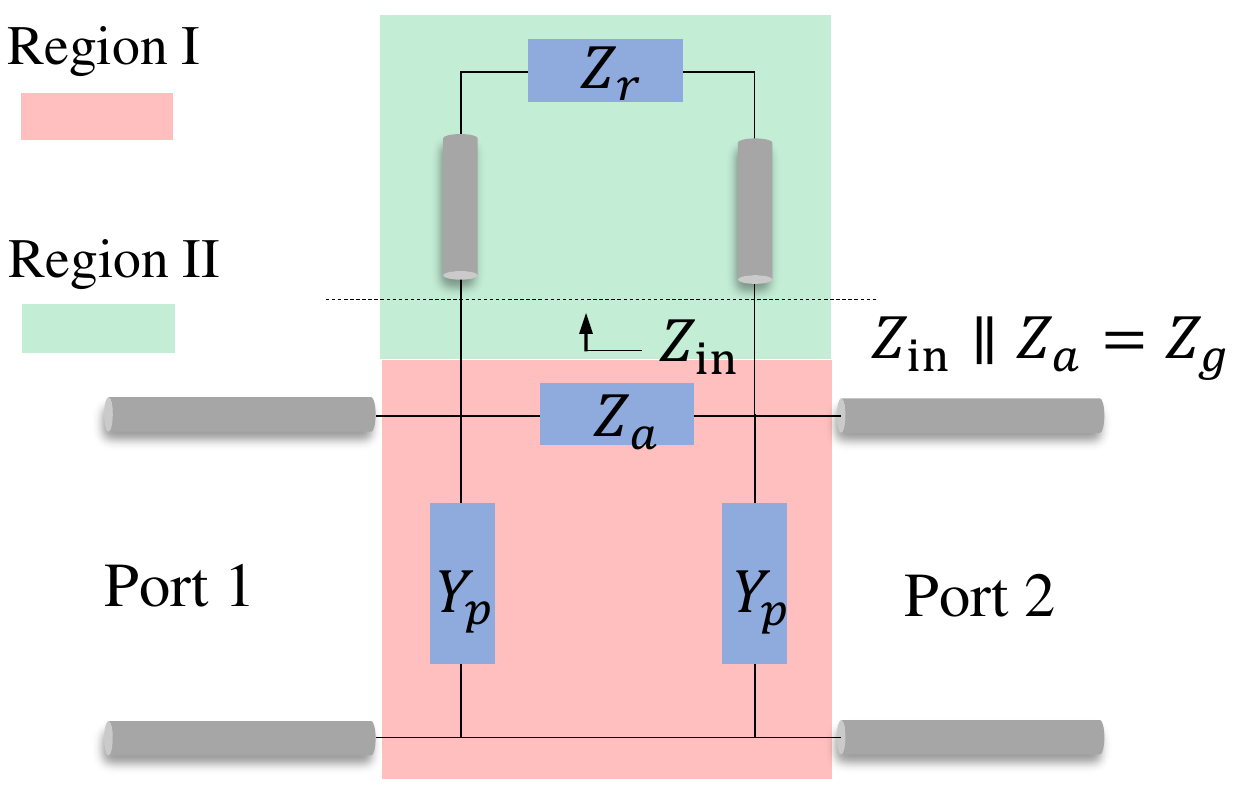}
		\caption{ Equivalent circuit of the measurement setup.  }\label{fig: equivalent circuit}
	\end{figure}

	Any passive two-port electromagnetic system can be modeled by an equivalent  $\Pi$-circuit. We model Region I as a $\Pi$-circuit formed by two parallel admittances $Y_p$ (these  two admittances are identical due to the structural symmetry) and one series impedance $Z_a$, as shown in Fig.~\ref{fig: equivalent circuit}. 
	For electrically thin gaps, the shunt admittance is capacitive (as will be numerically confirmed in the next section), because the vertical electrical field of TM$_n$ modes, $\mathbf{E}_y$, represents capacitive coupling between the top and bottom edges of the waveguide apertures.
	Region~II ($\mathbf{E}_y=0$) is an open-ended parallel-plate waveguide, which can be considered as a section of a transmission line terminated with an effective load impedance $Z_r$ as a model of the open end  ($Z_r$ includes edge reactance and the radiation resistance of the open end of the PPWG).
	The input impedance of Region~II (seen from Region~I) is denoted as $Z_{\rm in}$ which is a shunt connected to $Z_a$, as shown in Fig.~\ref{fig: equivalent circuit}. 
	From Fig.~\ref{fig: equivalent circuit}, it is obvious that the whole gap (including Regions~I and II) can be characterized as a unified $\Pi$-circuit, where the two shunt admittances $Y_p$ are inherited from Region~I and the series impedance $Z_g$ is formed by a parallel connection of $Z_{\rm in}$ and $Z_a$, denoted as $Z_g=Z_a \parallel Z_{\rm in}$.
	
	If the sample size is larger than Region~I, modifications of the sample shape change the electromagnetic environment in Region~II and thus influence $Z_{\rm in}$.
	Moreover, for millimeter and sub-terahertz waveguides, the flanges walls are normally not planar.  
	Any additional passive structures in Region~II, e.g., fastening screws, choke grooves, tapped holes, and alignment pins can be viewed as additional loadings of the PPWG, and thus they also affect the value of  $Z_{\rm in}$. 
	As a consequence, the total series impedance may significantly vary when the shape and size of the dielectric sample are changed. Even different positioning of the same sample or different tightness of fastening  affects the series impedance.
	However, the shunt impedance of the discontinuity will not be affected by these structural details in Region~II, since it is only determined by the capacitive coupling of the waveguide walls in Region~I which is very stable once the sample area is larger than Region~I. 
	For this reason, we can leverage the stable shunt impedance to characterize slab samples with arbitrary shapes positioned between arbitrary flanges. 
	Note that, the stability of shunt impedance was noticed in our previous work \cite{wang2017accurate}, but at that time, we did not realize that it can be used for permittivity extraction.

	\subsection{Stability of the shunt impedance} \label{section: stability}
	In the equivalent circuit of Fig.~\ref{fig: equivalent circuit}, the values of $Z_{g}$ and $Y_{p}$ can be extracted from measured $S$-parameters.  We use the transfer matrix method. After expressing the circuit components $Y_{ p}$ and $Z_{ g}$ in terms of $ABCD$ matrices, the total transfer matrix of the discontinuity can be calculated as the  cascaded multiplication of them:
	\begin{equation}
		\begin{split}
			\begin{bmatrix}
				A&B\\
				C&D\\
			\end{bmatrix}=&
			\begin{bmatrix}
				1&0\\
				Y_{ p}&1\\
			\end{bmatrix}
			\begin{bmatrix}
				1&Z_{ g}\\
				0&1\\
			\end{bmatrix}
			\begin{bmatrix}
				1&0\\
				Y_{ p}&1\\
			\end{bmatrix}\\[2ex]
			=&
			\begin{bmatrix}
				1+{Z_{ g}}Y_{ p}&{Z_{ g}}\\
				{Z_{ g}}Y_{ p}^2+2Y_{ p}&1+{Z_{ g}}Y_{ p} \label{eqt:matrix}
			\end{bmatrix}.
		\end{split}
	\end{equation}
	The matrix elements, $A$, $B$, $C$, and $D$, can be expressed as functions of $S$-parameters \cite[\textsection~4.4]{pozar2009microwave}. Therefore, we can relate the circuit values with $S$-parameters. Parameter $Z_{ g}$  can be expressed as
	\begin{equation}
		{Z_{ g}}=B=Z_{ 0}\frac{(1+S_{ 11})^2-S_{ 21}^2}{2S_{ 21}}.\label{Eqt:Xg}
	\end{equation}
	Here, $Z_{ 0}=\omega\mu_0/{\sqrt{\omega^2\mu_0\epsilon_0-(\frac{\pi}{L})^2}}$ is the characteristic  impedance of the $\rm {TE}_{\rm 10}$ mode in the rectangular waveguide. Another equation can be written as
	\begin{equation}
		1+Z_{ g}Y_{ p}=A=\frac{1-S_{ 11}^2+S_{ 21}^2}{2S_{ 21}}.\label{Eqt: A}
	\end{equation}
	Solving $Y_{ p}$ from (\ref{Eqt:Xg}) and (\ref{Eqt: A}), we obtain
	\begin{equation}
		{Y_{ p}}=\frac{A-1}{Z_{ g}}=\frac{1-S_{ 11}-S_{ 21}}{Z_{ 0}(1+S_{ 11}+S_{ 21})}.\label{Eqt:Yp}
	\end{equation}
	
\begin{figure}[h!]
		\centering
		\includegraphics[width=0.99\linewidth]{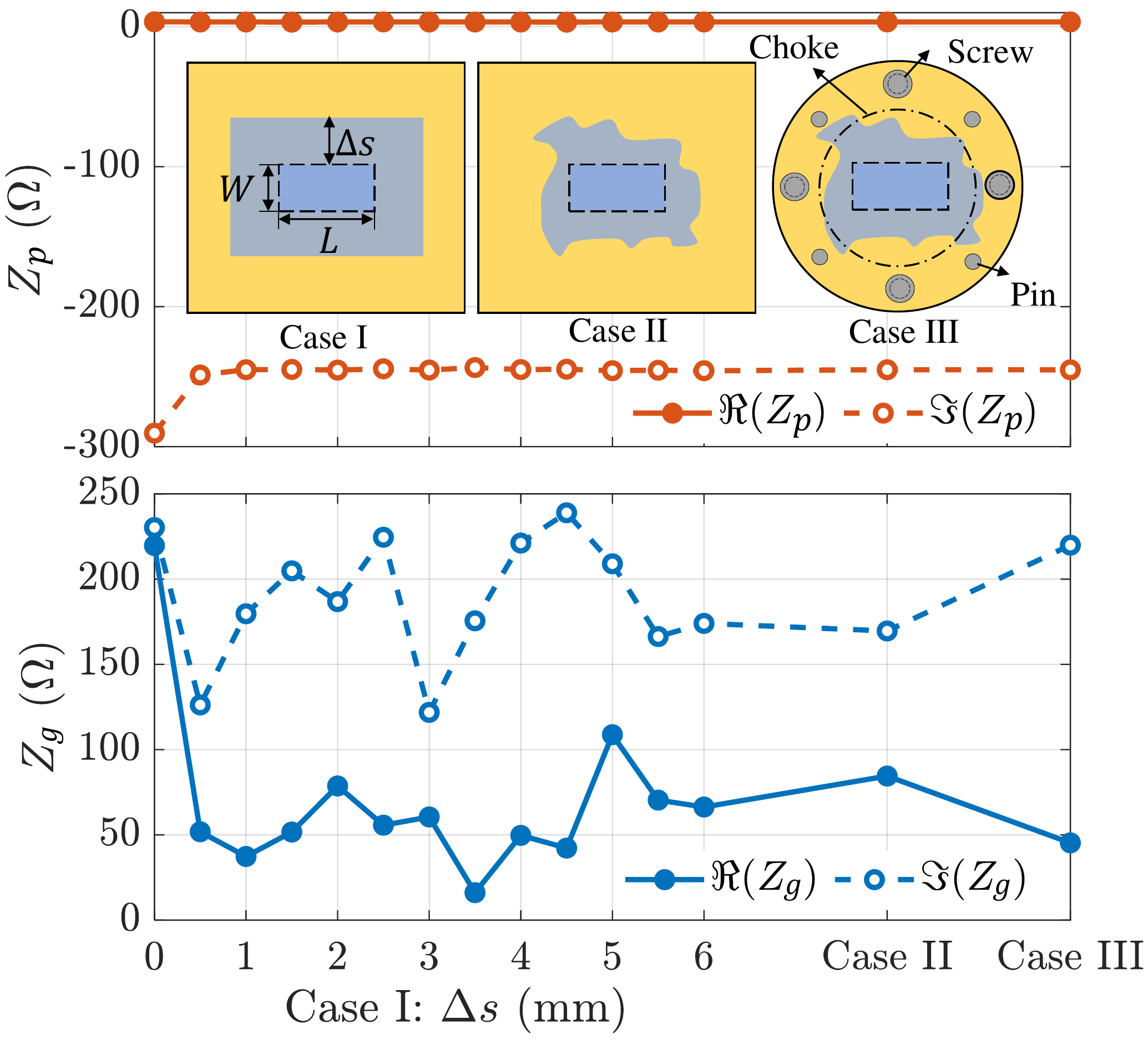}
		\caption{Extracted shunt (top) and  series (bottom) impedances for different sample dimensions.  In Case I, the sample size is $(L+\Delta s)\times(W+\Delta s)$. In Case II, the sample shape is randomly chosen but it is larger than the critical size. In Case III, the waveguide flanges are round with small accessory structures. }\label{fig:change_dimension}
	\end{figure}

	Next, we numerically demonstrate that $Y_{ p}$ is insensitive to the shape of the sample as well as to possible additional structures inside the waveguide discontinuity.
	In the simulations, the values of the permittivity and the thickness of the dielectric slab are the same as assumed in Fig.~\ref{fig: field model}(a). The measurement setup is modeled in three different ways. In the first case, the sample covers the waveguide aperture with dimension $W\times L$ and extends to the distance $\Delta s$ from the aperture edges, as illustrated in the first inset picture of Fig.~\ref{fig:change_dimension} (top). 
	We increase the extended size $\Delta s$ and extract the shunt impedance $Z_{ p}=1/Y_{ p}$ and series impedance $Z_{ g}$  from the simulated $S_{21}$ and $S_{11}$ according to Eqs.~(\ref{Eqt:Xg}) and (\ref{Eqt:Yp}). 
	It can be seen that as $\Delta s$ increases, the shunt impedance remains constant, while the series impedance is very unstable. This is because changes of the dielectric sample sizes modify the input impedance of Region~II and thus change $Z_{ g}$ dramatically. 
	Notice that when the sample size is close to the waveguide aperture ($\Delta s \approx0$), the extracted shunt impedance becomes sensitive to the size variations. This is because in this case the dominating $\rm TM_1$ mode does not fully decay in the sample, and variations of the sample size affect the field distribution in Region~I and therefore change the shunt impedance. 
	Obviously, there exists a  critical extension size $\Delta s_{\rm cr}$, for which the amplitude of the $\rm TM_1$ mode decays to $e^{-\alpha}$ ($\alpha$ is the decaying factor)  of the originally excited amplitude when it propagates in PPWG. According to this criterion,  $\Delta s_{\rm cr}$ can be calculated as
	\begin{equation}
		\Delta s_{\rm cr}=\frac{\alpha}{\sqrt{(\frac{\pi}{d})^2-\omega^2\epsilon_{\rm r}\epsilon_0\mu_0}}
	\end{equation}
	Therefore, the size of the test slab should be larger than $(L+\Delta s_{\rm cr})\times(W+\Delta s_{\rm cr})$ to ensure that the higher-order modes are negligible at the edges of the sample.
	The critical extension size for Case~I in Fig.~\ref{fig:change_dimension} is $\Delta s_{\rm cr}=0.69$~mm for $\alpha=3$. As we can see from Fig.~\ref{fig:change_dimension} (top panel), the shunt impedance of the gap does not change when $\Delta s>\Delta s_{\rm cr}$.
	
	In Case~II, we choose an arbitrarily shaped dielectric slab that is larger than the critical size. It is shown that the shunt impedance still keeps unchanged.  In Case III, the waveguide flanges are modified into a circular shape with actual screws, alignment pins, and choke grooves. We see that even with such a complicated gap environment, the shunt impedance $Y_{ p}$ is still not affected at all. 
	
	The above numerical experiments fully verify the predictions based on circuit modeling, confirming that the shunt impedance of the gap is insensitive to the gap environment as well as to  the shape and size of the sample. In the next section, we will show how to extract the permittivity from the measured $Y_{  p}$.

	\section{Characterization of electrically thin dielectric slabs}
	
	Next, we discuss how $Y_{ p}$ is related to the permittivity of the dielectric slab under test. Unlike the conventional waveguide characterization, here, the $S$-parameters have no explicit analytical relations with $\epsilon_{ r}$. Therefore, it is not straightforward to find the permittivity from measured $S$-parameters and the shunt admittance  $Y_{ p}$. 
	In this section, we discuss the extraction methods for electrically thin dielectric layers ($d<\lambda_{ d}/10$),  derive the extraction formulas, and show the measurement results.  
	
	\subsection{Extraction formula} \label{sec: extraction formula}
	
	For samples with ultra-subwavelength electrical thickness ($d<\lambda_d/10$), 
	the fields in Region~I are similar to the field in a material slab placed in a continuous rectangular waveguide. This is because very thin flange gaps have very large parallel-plate capacitances, allowing the currents on metal walls of the waveguides to pass through the gap.
	Therefore, the shunt impedance of Region~I can be approximated considering the same dielectric slab in a closed waveguide.  To find the shunt impedance of a dielectric slab inside the waveguide, a convenient way is to use the corresponding $ABCD$ matrix. By equating the matrix elements with that in Eq.~(\ref{eqt:matrix}), we can solve all the circuit components ($Y_{ p}$ and $Z_{ g}$).
	The $ABCD$ matrix of a dielectric slab inside a continuous waveguide can be expressed as
	\begin{equation}
		\begin{bmatrix}
			A&B\\
			C&D\\
		\end{bmatrix}=
		\begin{bmatrix}
			\cos{(\beta_{ d}d)}&jZ_{ d}\sin{(\beta_{ d}d)}\\
			jY_{ d}\sin{(\beta_{ d}d)}&	\cos{(\beta_{ d}d)}\\
		\end{bmatrix},\label{eqt:matrix1}
	\end{equation}
	where $\beta_{ d}=\sqrt{\omega^2\mu_0\epsilon_0\epsilon_{ r}-(\frac{\pi}{L})^2}$  is the  propagation constant in dielectric slab (TE$_{01}$ mode), and  $Z_{ d}=1/Y_{ d}=\mu_0\omega/\beta_{ d}$ is the corresponding characteristic  impedance. 
	After equating the matrix elements in Eq.~(\ref{eqt:matrix1}) and Eq.~(\ref{eqt:matrix}),  $Y_{ p}$ can be analytically solved:
	\begin{equation}
		Y_{ p}=\frac{\cos{(\beta_{ d}d})-1}{jZ_{ d}\sin(\beta_{ d}d)}=\frac{1-S_{ 11}-S_{ 21}}{Z_{ 0}(1+S_{ 11}+S_{ 21})}.\label{Eqt:7}
	\end{equation}
	Once the $S$-parameters are measured, the above equation  uniquely determines $\epsilon_{ r}$ [note that in Eq.~(\ref{Eqt:7}), $\beta_{ d}$ is a function of $\epsilon_{ r}$].
	
\begin{figure}[h]
		\centering
		\subfigure[]{\includegraphics[width=0.95\linewidth]{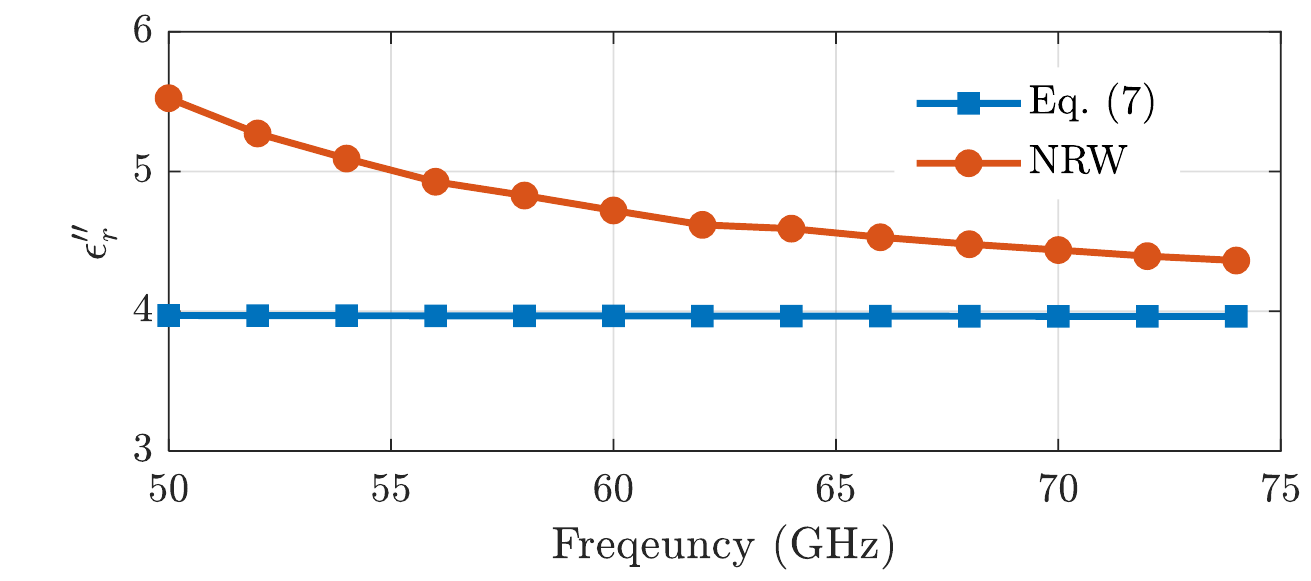}
			\label{fig:extraction_comparision_a}}
		\subfigure[]{\includegraphics[width=0.95\linewidth]{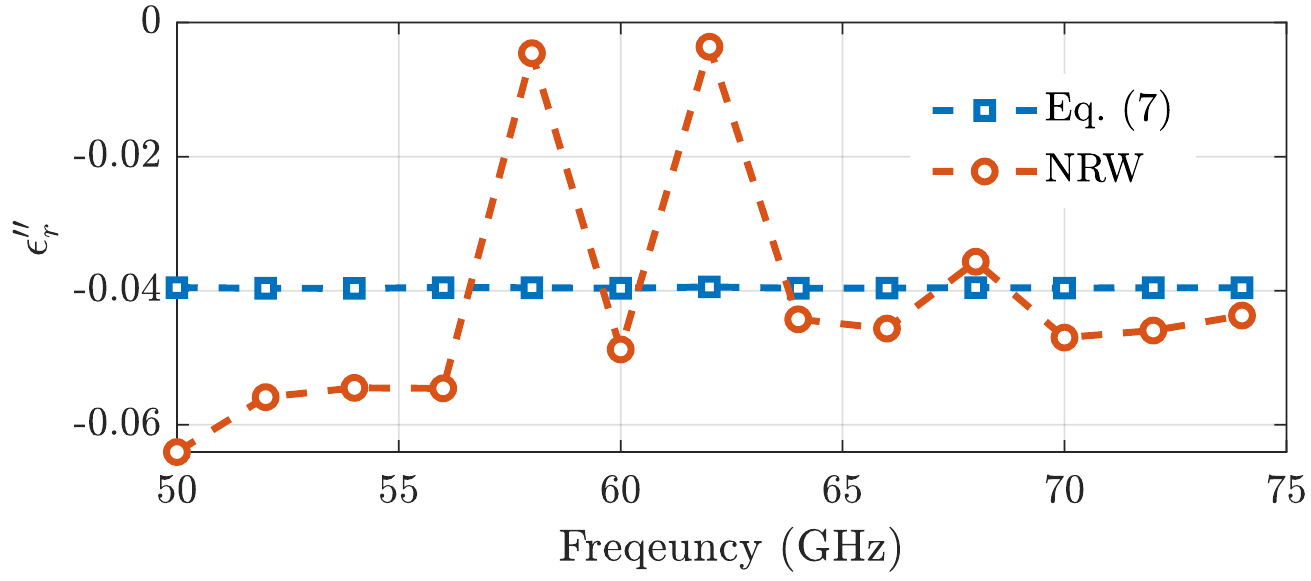}
			\label{fig:extraction_comparision_b}}
		\caption{ (a) Real and (b) imaginary (bottom) parts of permittivity solved from Eq.~(7), and using Nicolson–Ross–Weir (NRW) method. In the simulation setup, the sample is rectangular shaped with $\Delta_s=1$~mm. }\label{fig:extraction_comparision}
	\end{figure}

	To examine the accuracy of extraction formula  Eq.~(\ref{Eqt:7}),
	let us  consider a dielectric slab with   $d=100~\mu {\rm m}$ and $\epsilon_{ r}=\epsilon_{r}^\prime-j\epsilon_{ r}^{\prime\prime}=4-j0.04$. The electrical thickness is $d=\lambda_{ d}/25$ at 60~GHz.
	We simulate the setup with these assumed physical parameters and obtain the $S$-parameters from 50~GHz to 75~GHz. Using Eq.~(\ref{Eqt:7}), we solve the complex permittivity at each frequency point. The results are shown in Fig.~\ref{fig:extraction_comparision}. 
	The retrieved permittivity perfectly agrees with the value assumed in the simulation.
	It should be noted that although for thin slabs only a very small amount of power leaks away from the discontinuity, one cannot ignore it and use the conventional Nicolson-Ross-Weir formulas (which are derived for closed waveguides) to extract the permittivity. In Fig.~\ref{fig:extraction_comparision}, the extraction results using Eq.~(\ref{Eqt:7}) and the Nicolson-Ross-Weir formulas \cite{nicolson1970measurement}  are compared. We see  that even for such a thin gap, the  Nicolson-Ross-Weir method does not work due to the negligence of leaked power. 
	The proposed method is, however, fully applicable,  because it extracts the permittivity via the shunt impedance, but not directly from the $S$-parameters. 

	It is important to stress that the permittivity extraction formula  Eq.~(\ref{Eqt:7}) is only accurate for ultra-thin dielectric materials, i.e., when $|\beta_{ d}|d\ll 1$. Under this condition, we can make the following additional approximations in Eq.~(\ref{Eqt:7}): $\cos{(\beta_{ d}d})-1\rightarrow -(\beta_{ d}d)^2/2$ and $\sin{(\beta_{ d}d})\rightarrow \beta_{ d}d$, and obtain
	\begin{equation}
		Y_{ p}\approx\frac{j}{2}\left(\omega\epsilon_0\epsilon_{ r} -\frac{\pi^2}{\omega\mu_0L^2}\right)d. \label{eq: approx}
	\end{equation}
	Equation (\ref{eq: approx}) implies that, at a fixed frequency,  $\Re(Y_{ p})$ and $\Im(Y_{ p})$ are linearly dependent on $\epsilon_{ r}^{\prime\prime}$ and $\epsilon_{ r}^{\prime}$, respectively.

	\subsection{Measurement results}
	The extraction formula Eq.~(\ref{Eqt:7}) is suitable for thin-film characterization at microwave and millimeter-wave frequencies, e.g., Polyethylene Naphthalate (PEN) and Polyethylene Terephthalate (PET) films with the thickness  around one hundred microns which is much smaller than the wavelength. It is important to mention that measuring extremely thin sheets (several tens of microns) requires more accurate mechanical contact between the sample and flange walls. With loose fastening,  imperfection of  contact can be a noticeable error source.
	To avoid this problem, one can stack several layers of thin film to increase the thickness of the measured sample,  but the total thickness  still should  be much smaller than the wavelength.
	
	Here, we measure the permittivity of common copy paper and experimentally show the stability of shunt impedance. We stack four layers of $80~\mu$m thick STAPLES copy paper, forming a $320~\mu$m thick  sample ($d=\lambda_d/10$ at 60~GHz).
	The sample is cut into an arbitrary shape but larger than  $(L+\Delta s_{\rm cr})\times(W+\Delta s_{\rm cr})$ to ensure the stability of shunt impedance. The measurement comprises several steps:
	\begin{figure}[h!]
		\centering
		\includegraphics[width=0.8\linewidth]{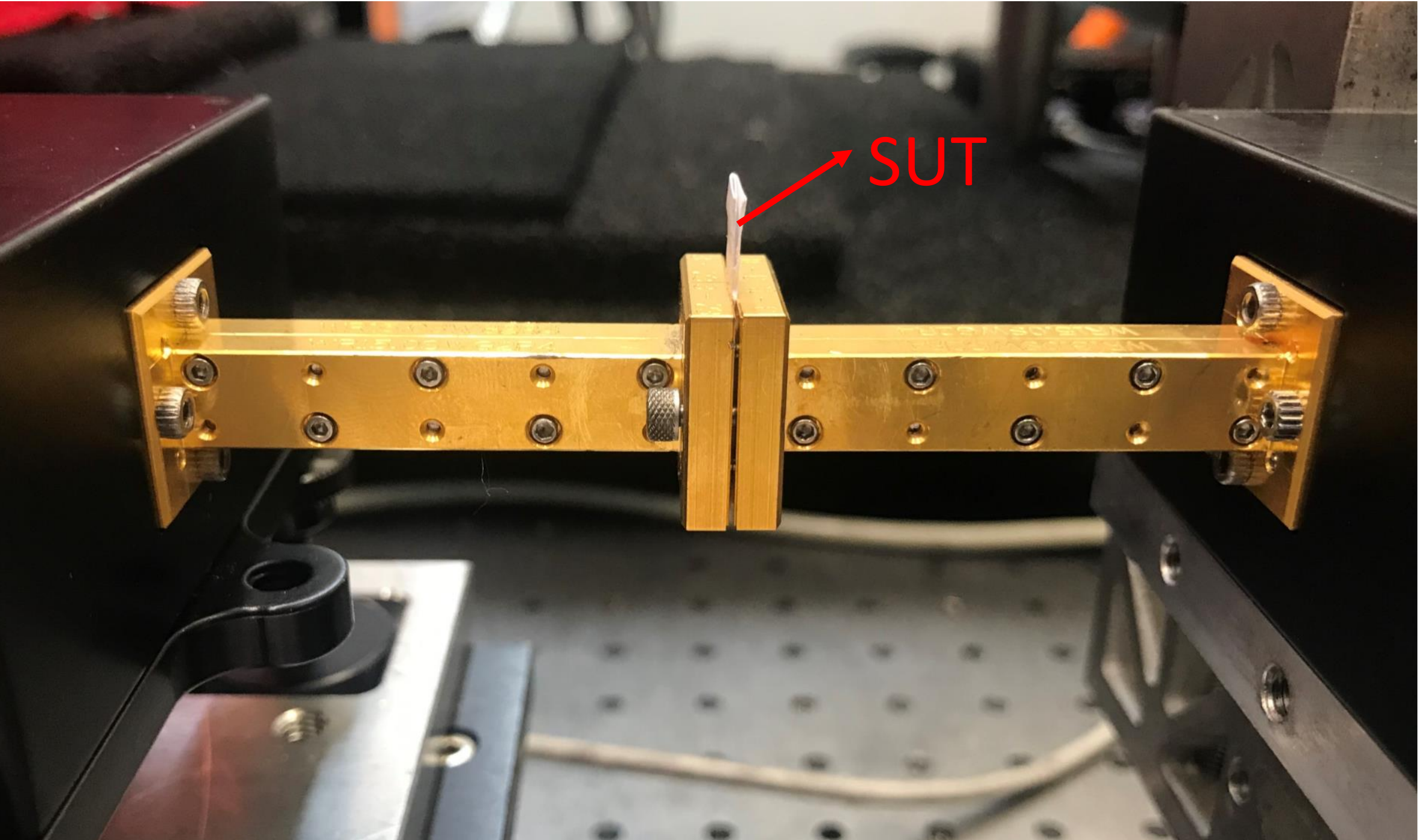}
		\caption{Photo of the measurement setup. In this example, the SUT is four layers of stacked copy papers. The waveguides are connected to WR-15 Vector Network Analyzer Extender (black modules).}\label{fig:photo}
	\end{figure}
	\begin{enumerate}
		\item Calibrate the system using  Thru-Reflect-Line (TRL) method. 
		\item Embed the sample between the flanges (see Fig.~\ref{fig:photo}). Note that it is not necessary to use a sample holder. One can cut the sample into a long strip that covers the waveguide aperture and hold it by hand when connecting the waveguides. After the sample was placed, close the waveguides and fasten the flanges using screws.
		\item Measure the $S$-parameters of the setup. At this step,  proper time gating can be applied to filter parasitic reflections caused by  waveguide misalignments. One should be careful not to remove the harmless reflections from the sample edge and other structures inside the flanges, otherwise, the extraction results will instead become less accurate. 
		
		\begin{figure}[h!]
			\centering
			\includegraphics[width=0.8\linewidth]{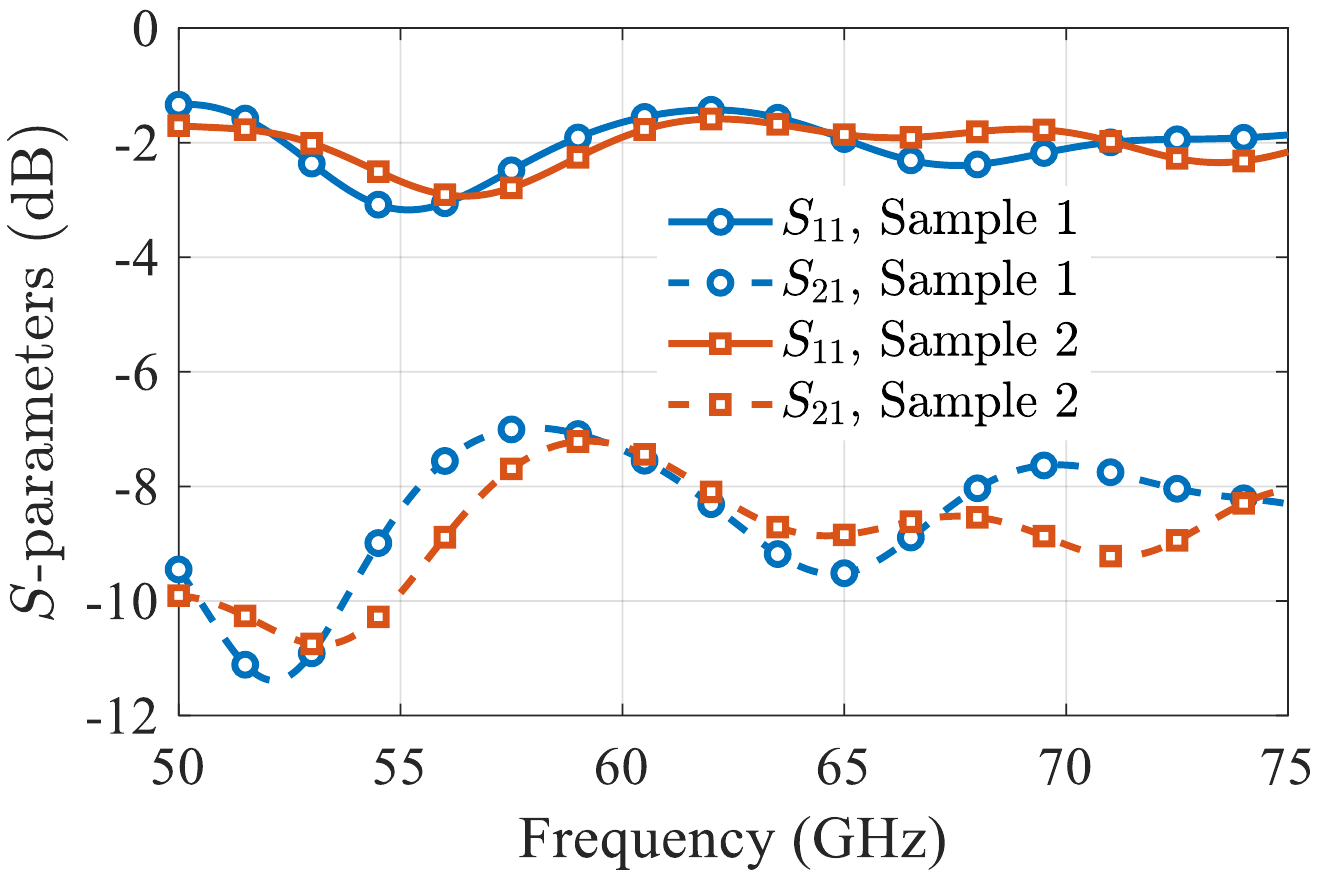}
			\caption{Measured $S$-parameters of  two paper samples of different shapes. }\label{fig:paper_S}
		\end{figure}

		\item Record the $S$-parameters and use Eq.~(\ref{Eqt:7}) to numerically extract the permittivity. 
	\end{enumerate}
	Figure~\ref{fig:paper_S} shows the measured magnitudes of $S$-parameters for two samples cut in arbitrarily different shapes. One can see that the measured $S$-parameters are obviously not the same for the two samples, since the sample shapes and sizes  in Region II are different.
	The difference in the measured $S$-parameters will pass on to the extracted series impedance, which is different for different samples, as shown in  Fig.~\ref{fig:Paper_impedance_re}. In contrast, the measured shunt impedance is very stable, as shown in  Fig.~\ref{fig:Paper_impedance_im}.
	
	The extracted complex permittivity [using Eq.~(\ref{Eqt:7})] of two paper samples is  shown in Fig.~\ref{fig:thin_measurement}. The results for the two samples are very close and stable in the studied frequency ranges, also agreeing with previously reported results  \cite{metaxas1974comparison}.  The slight difference might result from different actual thickness of the paper samples caused by different tightness of the screws. 
	\begin{figure}[h]
		\centering
		\subfigure[]{\includegraphics[width=0.9\linewidth]{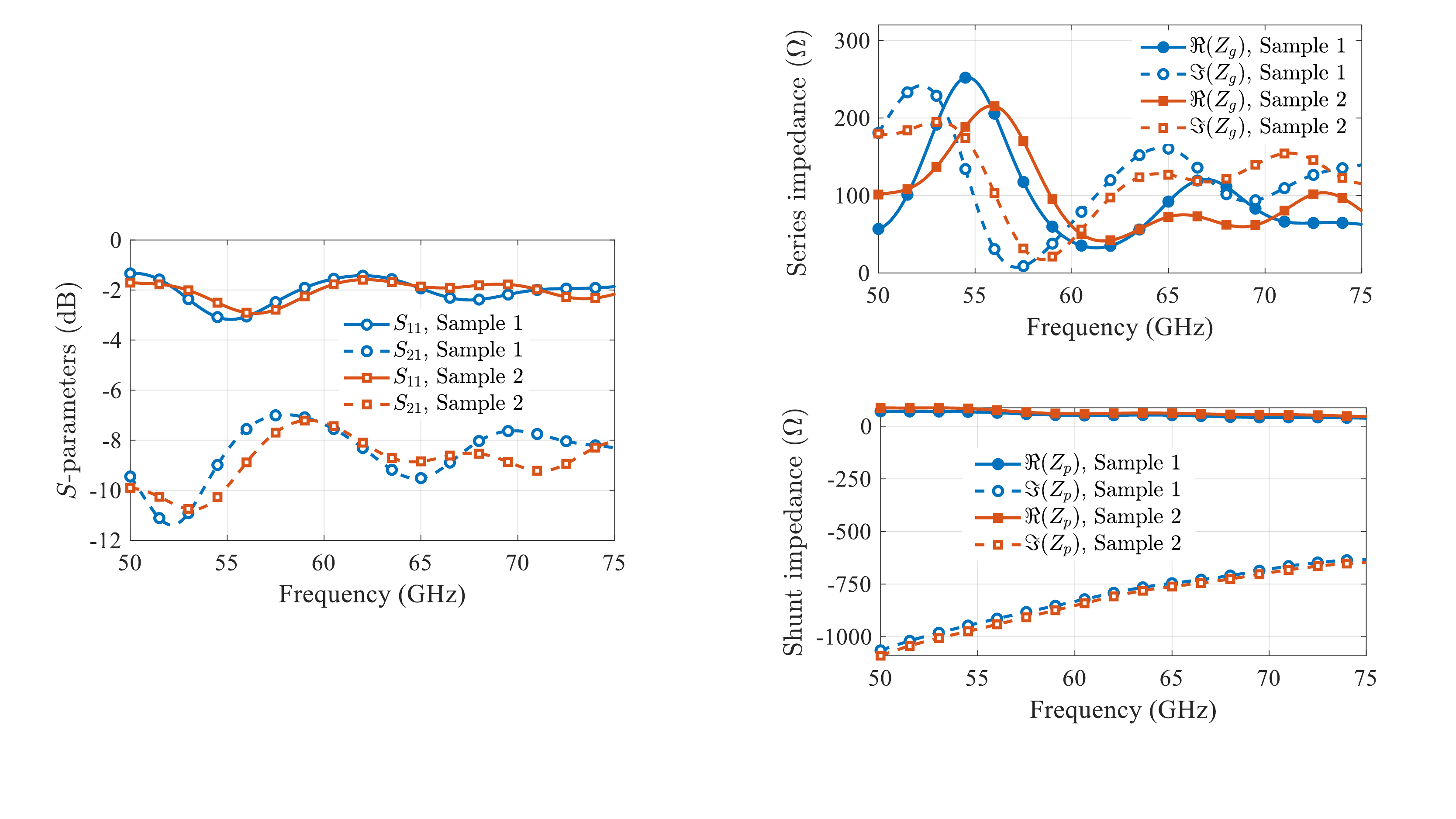}
			\label{fig:Paper_impedance_re}}
		\subfigure[]{\includegraphics[width=0.9\linewidth]{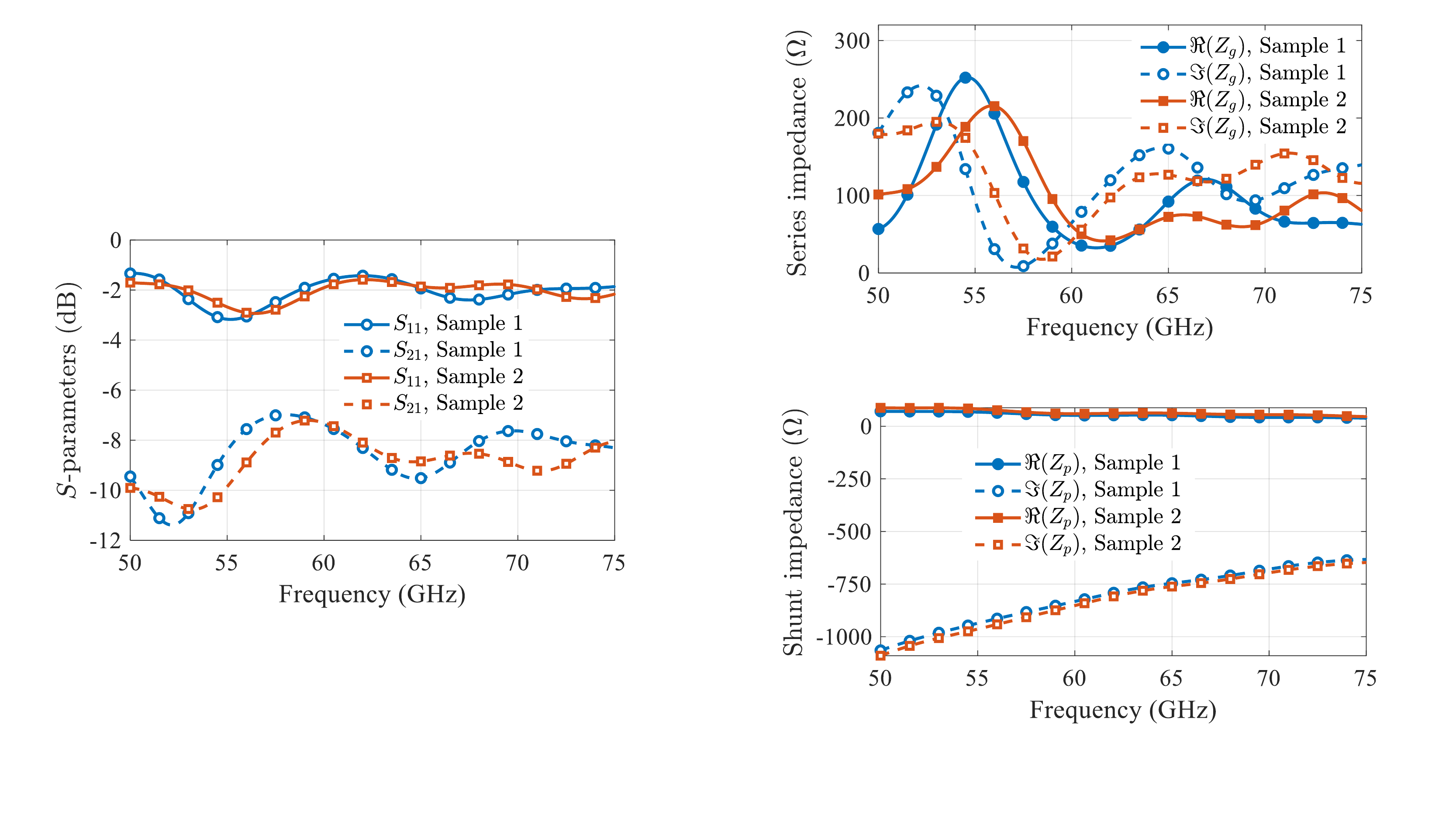}
			\label{fig:Paper_impedance_im}}
		\caption{ Extracted (a) series  and (b) shunt  impedances of the discontinuity created by two different samples. }\label{fig:thin_measurement}
	\end{figure}
	
	To further verify the accuracy of the method, we measure the permittivity of Polyethylene Naphthalate (PEN) layers and, as a further validation check, of free space. 
	PEN samples are stacked in two layers (the thickness of each layer is $125~\mu$m), and the total thickness is about $d\approx\lambda_{ d}/12$ at 60~GHz. 
	The extracted permittivity is shown in the yellow curves in Fig.~\ref{fig:thin_measurement}. The measured value is around $\epsilon_{ r}=3.05-j0.05$, being in good agreement with the previously reported values in \cite{bisognin2014inkjet,khanal2016towards} at millimeter-wave frequencies. We also measure the permittivity of air. The ``air sample'' is formed by opening an arbitrary shaped (but larger than the critical size) hole in an $400~\mu$m thick FR4 laminates. The extraction results are shown in  Fig.~\ref{fig:thin_measurement} (purple curves), confirming  good accuracy.
	In all the measured samples, it appears that the imaginary part of permittivity suffers more perturbations than the real part. This is caused by  uncertainties of measured $S$-parameters. The reason will be explained Sec.~\ref{Sec: uncertainty} where the measurement uncertainties for both real and imaginary parts of permittivity are analyzed in detail.

	Importantly, one should remember that the extraction formula Eq.~(\ref{Eqt:7}) is only accurate for electrically thin materials ($d<\lambda_d/10$). As the electrical thickness of SUT increases, the extraction formula gradually becomes inaccurate. 
	
	\begin{figure}[h]
		\centering
		\subfigure[]{\includegraphics[width=0.9\linewidth]{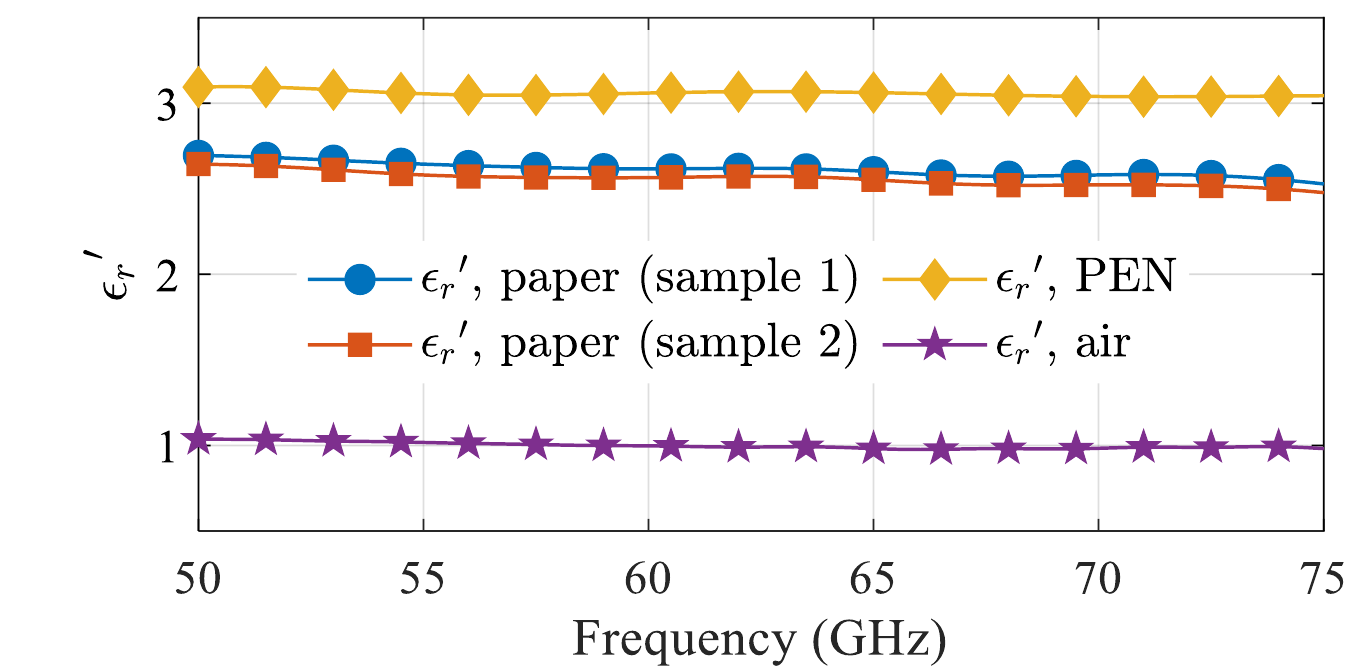}
			\label{fig:thin_measurement_re}}
		\subfigure[]{\includegraphics[width=0.9\linewidth]{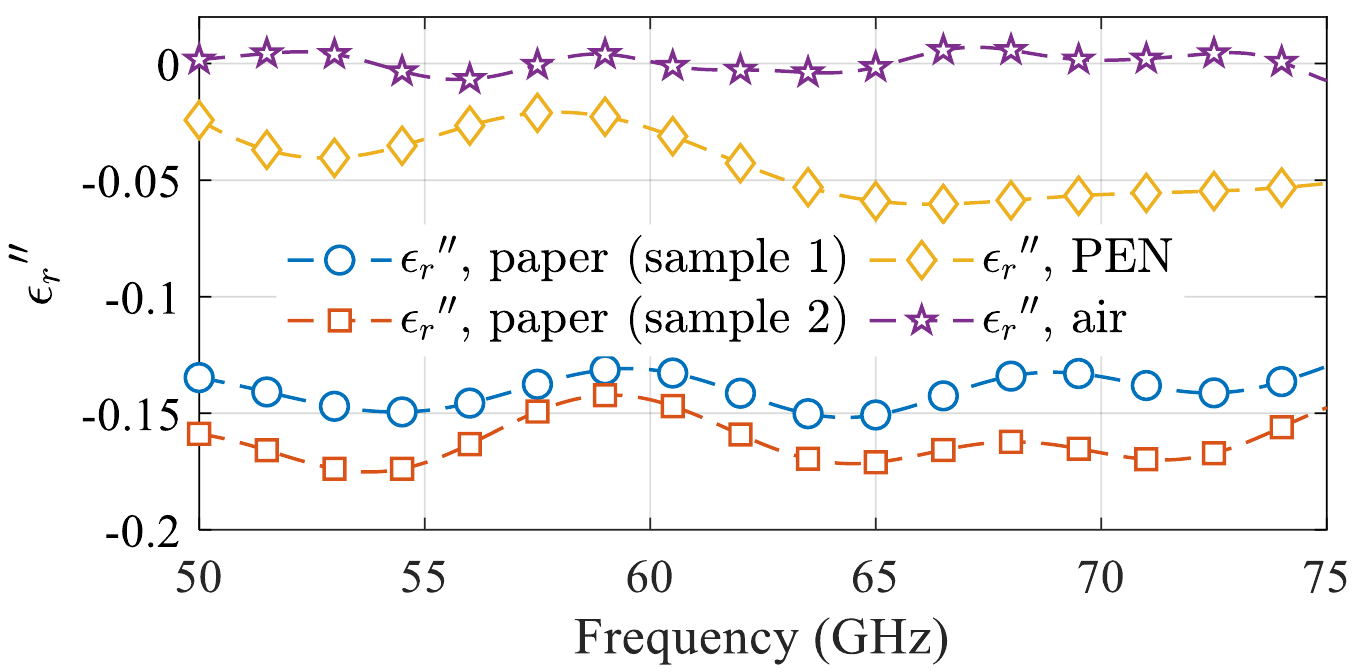}
			\label{fig:thin_measurement_im}}
		\caption{  Extracted (a) real and (b) imaginary parts of permittivity for different types of materials.  }\label{fig:thin_measurement}
	\end{figure}

	\section{Characterization of thick dielectric layers} \label{sec: thick}
	
	When the thickness of dielectric slabs increases ($\lambda_{ d}/10<d<\lambda_{ d}/2$), the  higher-order TM modes become more and more significant  in Region I, and the field in Region I can be obviously different from the field in the closed waveguide. Therefore, one cannot use a simple transmission-line section model for Region I, and the extraction formula  Eq.~(\ref{Eqt:7}) becomes inaccurate. Obviously, the relation between $Y_{ p}$ and $\epsilon_{ r}$ is not as straightforward as for thin samples.

	\subsection{Simulation-assisted extraction method} \label{simulationassisted}
	
	Here, we utilize numerical tools (Ansys HFSS), to find the relation between $Y_{ p}$ and $\epsilon_{ r}$. Numerical fitting is a common method to extract material parameters from measured data.  By modeling the measurement setup in numerical tools and fitting the simulated $S$-parameters with the measured values, one can estimate the permittivity of the sample. 
	In the conventional numerical fitting method, one should accurately model the actual measurement setup \cite{nefedova2015dielectric} since the modeling errors can induce significant inaccuracy in the simulated $S$-parameters and thus result in erroneous estimations of permittivity.  
	To overcome this problem, instead of fitting the scattering parameters, we fit the equivalent shunt impedance/admittance. As we demonstrated in Sec.~\ref{section: stability}, the shunt impedance/admittance is only related to the thickness (which is easy to measure) and the sample permittivity, and is not affected by details of the discontinuity and external environment.  In this way, one can avoid the need to accurately reproduce all the setup details in simulation tools. The measurement procedure compromises the following steps: 
	
	\begin{enumerate}
		\item The sample is measured in a waveguide junction.  The shunt admittance is extracted from the measured $S$-parameters using Eq.~(\ref{Eqt:Yp}). 
		\item The physical setup is modeled in the simulation tool. Note that it is not necessary to accurately model the measurement setup in the simulation since the complicated structures of the waveguide junction and the shape of the sample (as long as it is larger than the critical size) does not affect the shunt admittance. 
		\item At the frequencies of interest,  different values of permittivity (both real and imaginary parts) are assumed in the simulation,  and the shunt admittance is extracted from numerical results. The simulated shunt impedance is then compared with the measured values. When the simulated and measured values are identical, the permittivity assumed in the simulation is the actual permittivity of the material under test. 
	\end{enumerate}
	\begin{figure}[h]
		\centering
		\subfigure[]{\includegraphics[width=0.44\linewidth]{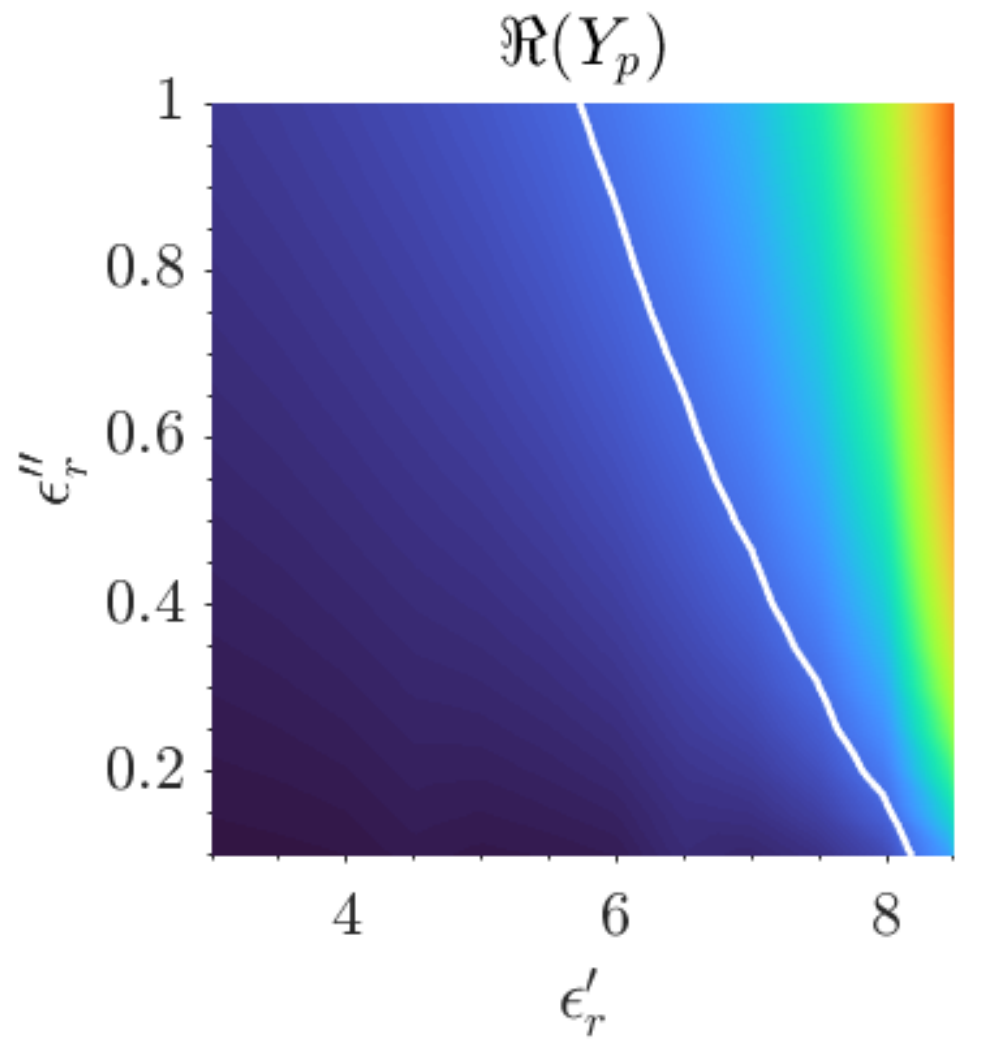}
			\label{fig:colormap_thick_re}}
		\subfigure[]{\includegraphics[width=0.51\linewidth]{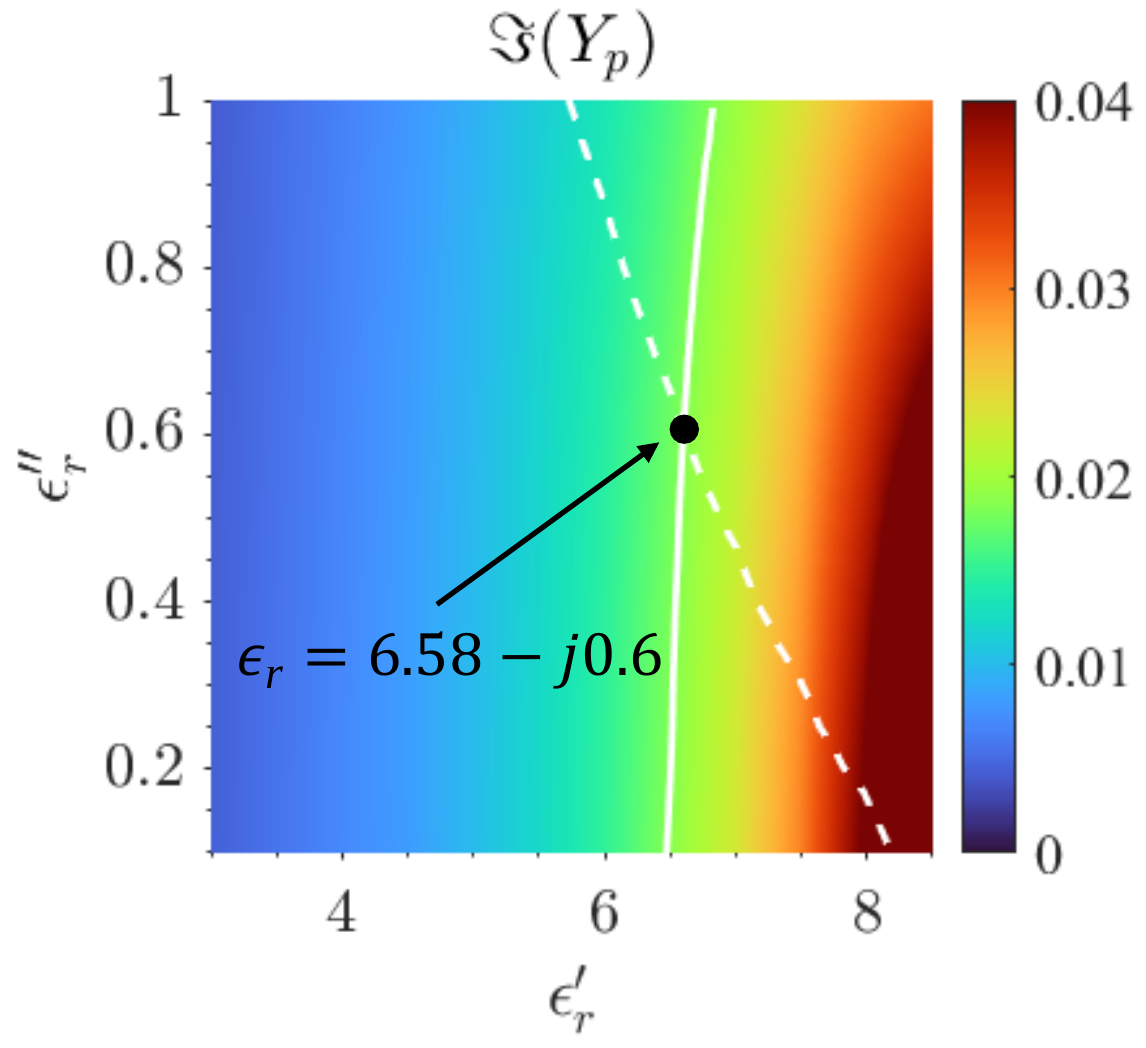}
			\label{fig:colormap_thick_im}}
		\caption{  (a) Real and (b) imaginary parts of shunt admittance at 75~GHz in terms of real and imaginary parts of permittivity. The data is obtained from numerical simulation using HFSS. }\label{fig:colormap}
	\end{figure}
	
	For example, we assume the thickness of SUT is $d=660~\mu$m and the frequency of interest is $f=75$~GHz.
	We model the measurement setup in HFSS and vary $\epsilon_{ r}^\prime$ and $\epsilon_{ r}^{\prime\prime}$ within reasonable ranges. For each pair of $\epsilon_{ r}^\prime$ and $\epsilon_{ r}^{\prime\prime}$, we can calculate the corresponding shunt admittance using Eq.~(\ref{Eqt:Yp}). In this way, we can plot  $\Re(Y_{ p})$ and $\Im(Y_{ p})$ as functions of $\epsilon_{ r}^\prime$ and $\epsilon_{ r}^{\prime\prime}$, as shown in Fig.~\ref{fig:colormap}. 
	The measured shunt admittance at 75~GHz is $Y_{ p}=0.005 + j0.018$.
	Then, we draw two contour curves $\Re(Y_{ p})=0.005$ and $\Im(Y_{ p})=0.018$ in Fig.~\ref{fig:colormap}(a) and (b) (white solid curves), respectively. 
	The intersection point of the two contour curves ($\epsilon_r=6.58-j0.6$) in Fig.~\ref{fig:colormap}(b) is the measured value of permittivity at 75~GHz.

	\subsection{Measurement results}
\begin{figure}[h]
		\centering
		\subfigure[]{\includegraphics[width=0.95\linewidth]{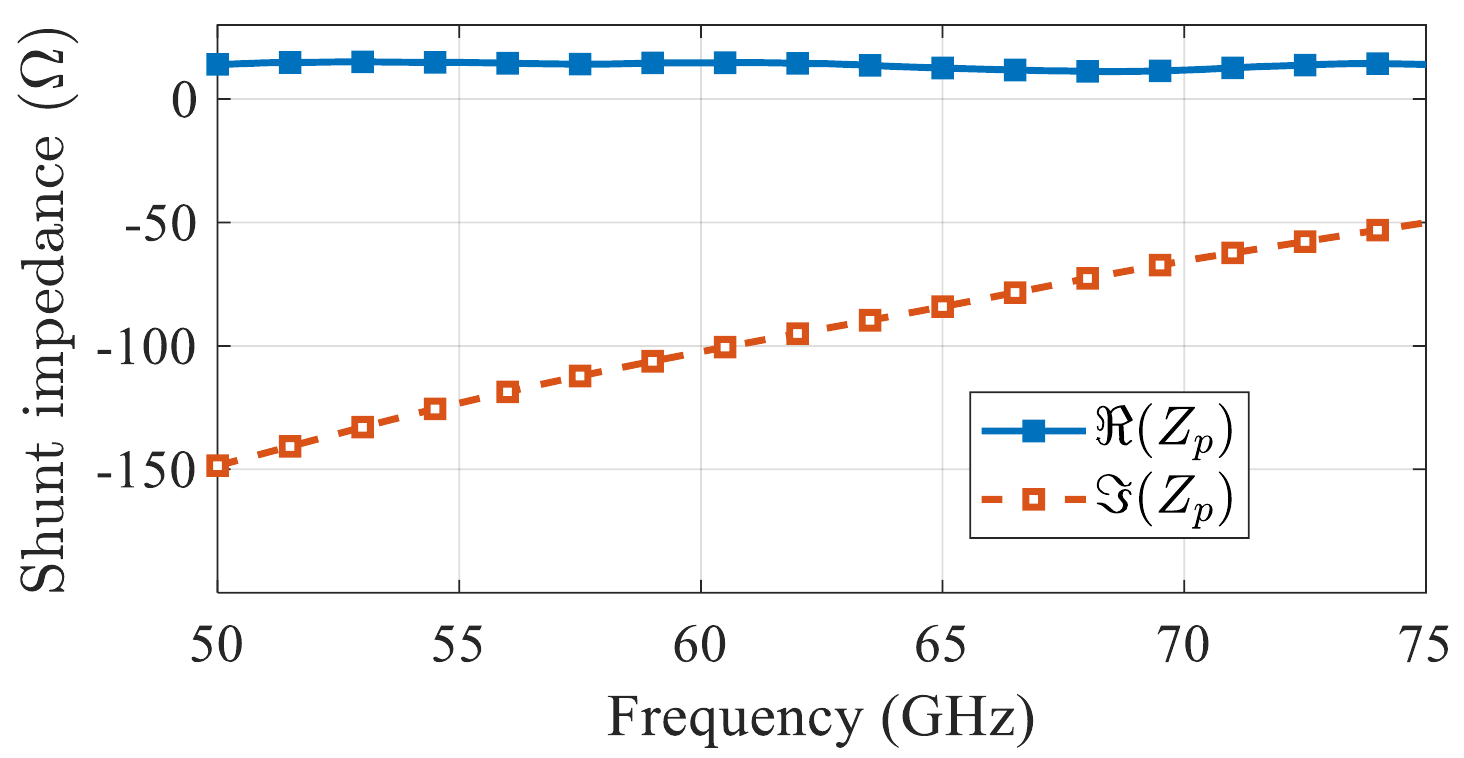}
			\label{fig:thick_measurement_a}}
		\subfigure[]{\includegraphics[width=0.95\linewidth]{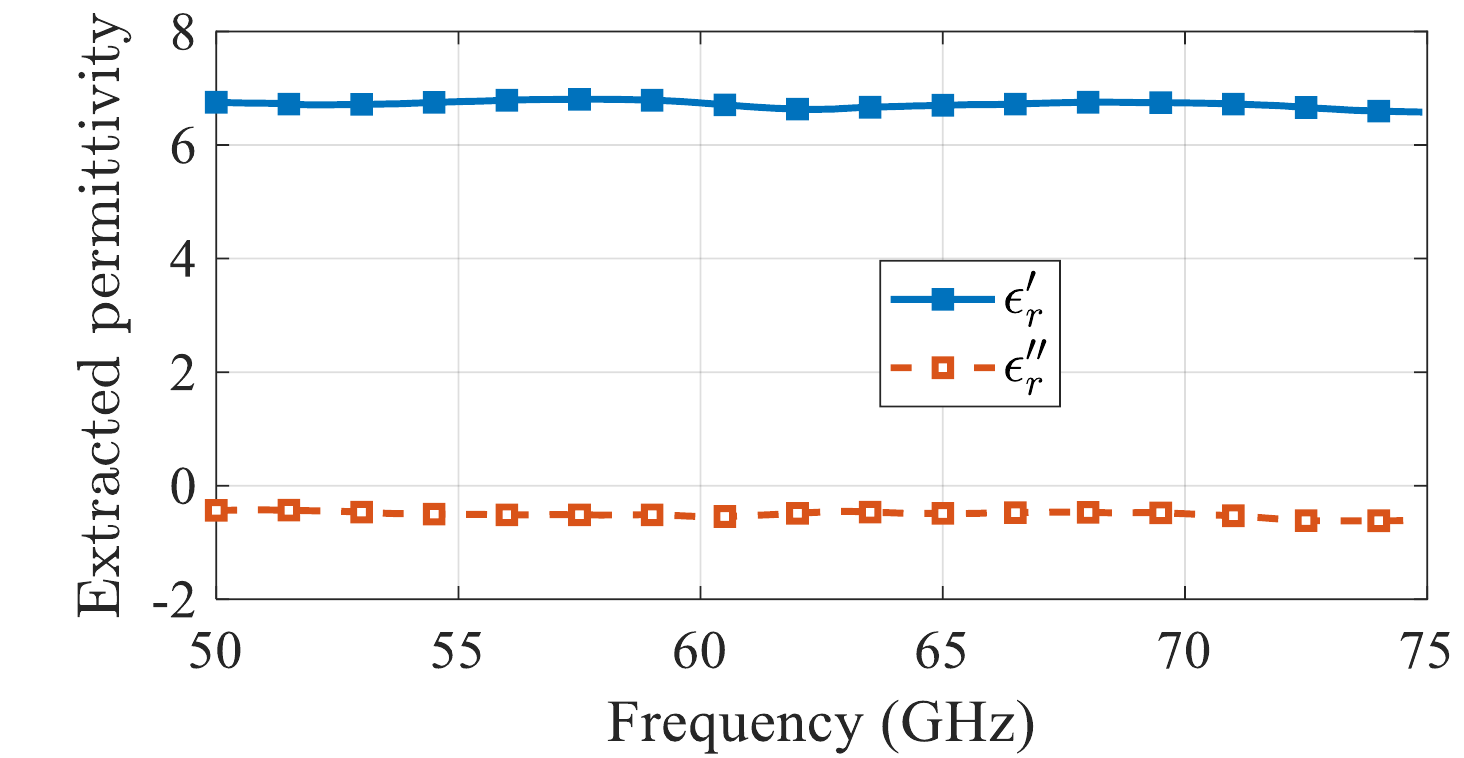}
			\label{fig:thick_measurement_b}}
		\caption{Measured shunt impedance (a) and  permittivity (b) of Corning$^{\circledR}$ Gorilla$^{\circledR}$ Glass 6 used for screens of mobile devices.  }\label{fig:thick_measurement}
	\end{figure}

	In this section, we measure the permittivity of a mobile phone  screen glass (Corning$^{\circledR}$ Gorilla$^{\circledR}$ Glass 6) with the thickness $d=660~\mu$m. The shunt admittance is extracted from the measured $S$-parameters, as shown in Fig.~\ref{fig:thick_measurement}(a).
	To extract the permittivity at all measured frequencies, it is not efficient to fit the permittivity value at each frequency one by one, following the procedure introduced at the end of Sec.~\ref{simulationassisted}. Here, we use a deep-learning technique to analyze the simulation data and quickly extract the permittivity at all the frequencies of interest.
	
	In numerical simulations, we model the setup and perform parametric studies in terms of $f$, $\epsilon_{ r}^{\prime}$, and $\epsilon_{ r}^{\prime\prime}$. 
	From simulations, we obtain more than 2000 sets of data, $[f,\epsilon_{ r}^{\prime}, \epsilon_{ r}^{\prime\prime}, \Re(Y_{ p}), \Im(Y_{ p})]_{\rm simu}$.
	The task is to use the simulated dataset to find $\epsilon_{ r}^{\prime}$ and $\epsilon_{ r}^{\prime\prime}$ for a given set of $[f, \Re(Y_{ p}), \Im(Y_{ p})]_{\rm meas}$ that is obtained from measurements. 
	This is a multi-dimensional fitting problem. 
	We use the Neural Net Fitting app in MATLAB to train a fitting network. 
	In the model training, the input datasets are  $[f, \Re(Y_{ p}), \Im(Y_{ p})]_{\rm simu}$ and the output datasets are $[\epsilon_{ r}^{\prime}, \epsilon_{ r}^{\prime\prime}]_{\rm simu}$. 
	The Levenberg-Marquardt Algorithm is chosen to train the neural network. For 2000 datasets, training can be completed within several seconds. Once the fitting model is trained,  the measured datasets $[f, \Re(Y_{ p}), \Im(Y_{ p})]_{ \rm meas}$ are fed to the model as inputs, and the output is the predicted permittivity. Figure~\ref{fig:thick_measurement}(b) shows the extracted permittivity in the measured frequency range. 
	
	The real part of permittivity is between 6.6 and 6.8, which is in good agreement with the reference value ($\epsilon_{ r}=6.69-j0.087$ at $f=3$~GHz, measured in \cite{corning}).  At millimeter-wave frequencies, the material loss (the measured values $0.4<\epsilon_{ r}^{\prime\prime}<0.66$ over this frequency range)  significantly increases as compared to the provided value at microwave frequencies.

	\section{Uncertainty Analysis} \label{Sec: uncertainty}
	
	The measurement uncertainty  originates from  inaccuracies in the measurement  of sample thickness, flanges alignments,  imperfect contact between SUT and flange walls, and so on. 
	The measurement errors caused by those factors can be reduced by using high-precision thickness characterization devices (e.g., profilometers) and careful assembling of the measurement setup.  
	Other important sources of measurement errors include  uncertainties of the measured $S$-parameters (both magnitude and phase), which are 
	unavoidable and determined by the  VNA device parameters. 
	
	In this section, we study the measurement errors caused by  uncertainties of the measured  $S$-parameters. 
	The  uncertainties of $S$-parameters on both magnitude and phases are denoted as $|\Delta S_{\alpha}|$ and $\Delta \theta _{\alpha}$ ($\alpha=11, 21$). 
	To analyze the impact of these parameters on the extracted permittivity, we use the differential method, where the dependent variable, $\epsilon_{ r}$, is differentiated with respect to each possible error parameter (the magnitudes and phases of the $S$-parameters involved in the extraction method) \cite{catala2003accurate,wang2017accurate}. 
	Since each derivative can take positive or negative values, the final error is calculated as a sum of the squared values of all derivatives:
	\begin{equation}
		\Delta \epsilon_{ r}=\sqrt{\sum_{\alpha}\left(\frac{\partial \epsilon_{ r}}{\partial |S_{\alpha}|} \Delta|S_{\alpha}|\right)^2+
			\sum_{\alpha}\left(\frac{\partial \epsilon_{ r}}{\partial \theta_{\alpha}} \Delta \theta_{\alpha}\right)^2},\label{uncertainty formula} 
	\end{equation}
	where
	\begin{equation}
		\frac{\partial \epsilon_{ r}}{\partial |S_{\alpha}|}= 
		\frac{ \partial \epsilon_{ r}}{\partial Y_{ p}} \frac{ \partial Y_{ p}}{\partial |S_{\alpha}|}, \quad
		\frac{\partial \epsilon_{ r}}{\partial \theta_{\alpha}}= 
		\frac{ \partial \epsilon_{ r}}{\partial Y_{ p}} \frac{\partial Y_{ p}}{\partial \theta _{\alpha}}.\label{eq: partial}
	\end{equation}
	In Eq.~(\ref{eq: partial}),
	\begin{equation}
		\frac{\partial Y_{ p}}{\partial |S_{\alpha}|}=\frac{-2}{Z_0F^2}\frac{S_{\alpha}}{|S_{\alpha}|}, \quad  \frac{\partial Y_{ p}}{\partial \theta _{\alpha}}=\frac{-2j}{Z_0F^2}S_{\alpha},
	\end{equation}
	with $F=1+S_{11}+S_{21}$. Furthermore, in Eq.~(\ref{eq: partial}), ${ \partial \epsilon_{ r}}/{\partial Y_{ p}}=2/\left(j\omega\epsilon_0 d\right)$ according to Eq.~(\ref{eq: approx}) for thin dielectric samples. For a thick layer, the linear dependence of $Y_{ p}$ on  $\epsilon_{ r}$  does not hold, as seen from Fig.~\ref{fig:colormap}. Therefore, numerical fitting techniques, e.g., `lsqnonlin' function in MATLAB, are needed to model the nonlinear relation between $\epsilon_{ r}$ and $Y_{ p}$. 
	
	\begin{figure}[h]
		\centering
		\subfigure[]{\includegraphics[width=0.95\linewidth]{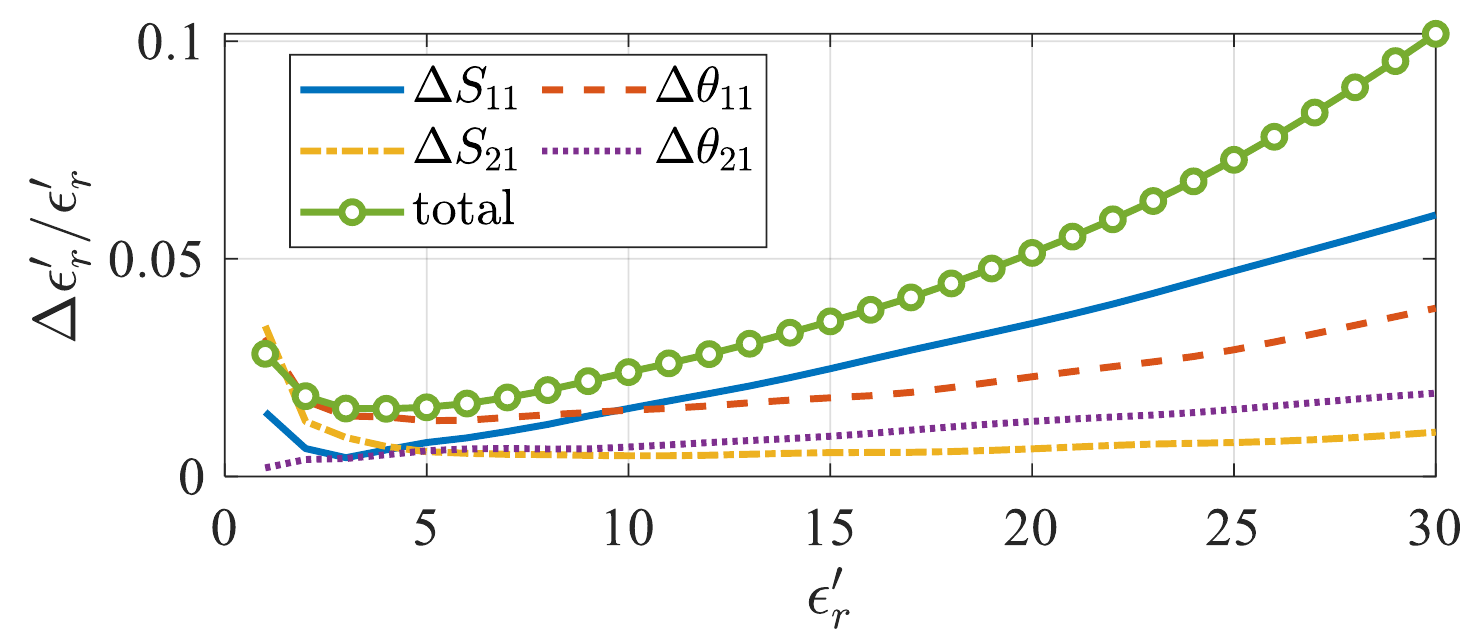}
			\label{fig:uncertainty_real}}
		\subfigure[]{\includegraphics[width=0.95\linewidth]{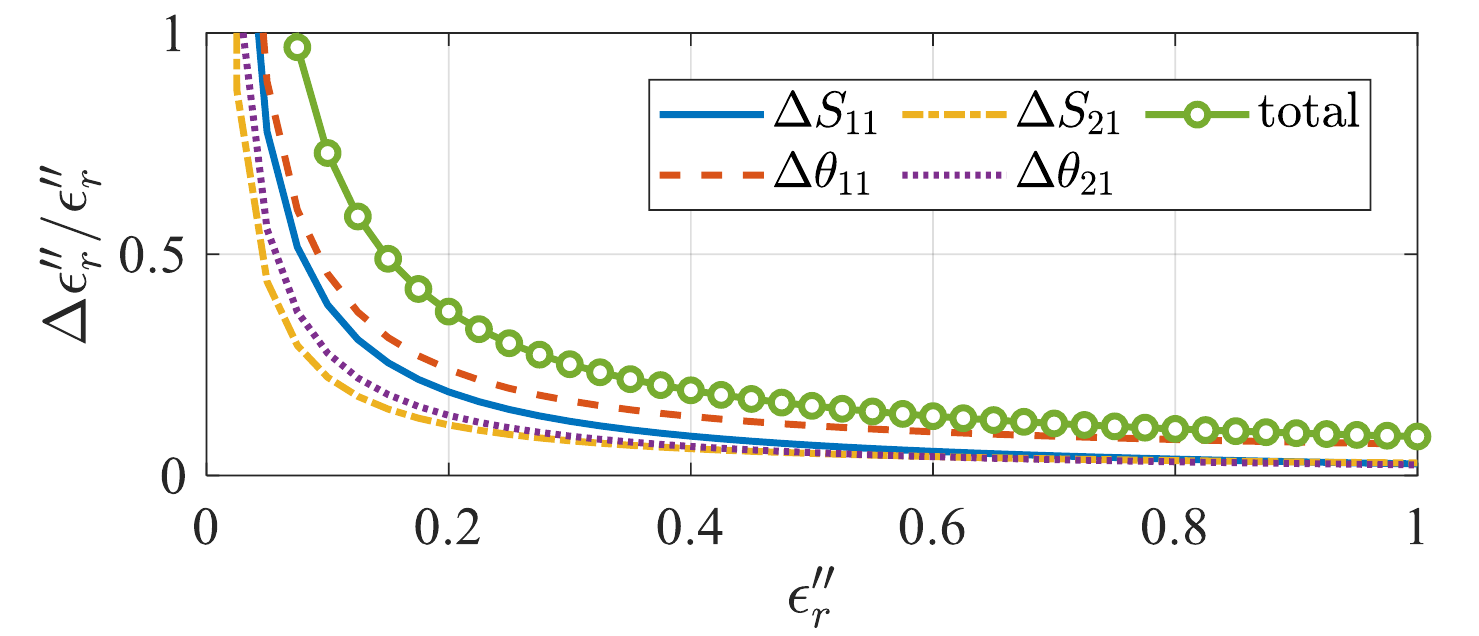}
			\label{fig:uncertainty_imag}}
		\caption{  Measurement uncertainties for (a) real and (b) imaginary parts of permittivity caused by each error source ($|\Delta S_{\alpha}|$ and $\theta _{\alpha}$) and their total effects.  }\label{fig:uncertainty}
	\end{figure}
	
	Let us assume that the sample thickness is $d=280~\mu$m and the testing frequency is $f=60$~GHz. In the first analysis, we evaluate the uncertainty of the real part of permittivity. To do this, we fix $\tan\delta=0.01$ and sweep $\epsilon_{ r}^\prime$ from $\epsilon_{ r}^\prime=1$ to $\epsilon_{ r}^\prime=30$ in simulation, and obtain the corresponding $S_{\alpha}$ for each permittivity value. 
	In real measurements, the uncertainties of $|\Delta S_{\alpha}|$ and $\Delta \theta _{\alpha}$ depend on the magnitude of $S_{\alpha}$, and the dependence can be obtained from Keysight Uncertainty Calculator for a specific vector analyzer \cite{keysight}. In this work, we choose E8361C Vector Network Analyser with V11644A Calibration Kit in the calculator. 
	Therefore,
	for each set of  $\epsilon_{ r}^\prime$ and  $\epsilon_{ r}^{\prime\prime}$, we can calculate the uncertainties of $\Delta\epsilon_r^\prime$ and $\Delta\epsilon_r^{\prime\prime}$  
	using Eq.~(\ref{uncertainty formula}), where ${\partial \epsilon_{ r}}/{\partial |S_{\alpha}|}$ and ${\partial \epsilon_{ r}}/{\partial |\theta_{\alpha}|}$ are obtained from the simulation, and,  $\Delta S_{\alpha}$  and $\Delta \theta_{\alpha}$ are provided by VNA manufacturer. 
	
	The uncertainty of $\epsilon_{ r}^\prime$ is plotted in Fig.~\ref{fig:uncertainty}(a). 
	It can be seen that the measurement uncertainty remains relatively low (below $5\%$) for $\epsilon_{ r}^\prime$ ranging from $\epsilon_{ r}^\prime=1$ to $\epsilon_{ r}^\prime=20$. 
	The uncertainty reaches its minimum ($1.5\%$) for $\epsilon_{ r}^\prime\approx4$. For larger $\epsilon_{ r}^\prime$, $\Delta \epsilon_{ r}^\prime$ increases. This is because as the slab becomes more reflective, the uncertainty of $|S_{ 11}|$ increases as its magnitude increases, which results in a decrease in the measurement accuracy. 
	
	In the second example analysis, the uncertainty of the imaginary permittivity is evaluated. In this case, we fix  $\epsilon_{ r}^\prime=5$ and vary $\epsilon_{ r}^{\prime\prime}$ from $\epsilon_{ r}^{\prime\prime}=0$ to $\epsilon_{ r}^{\prime\prime}=1$ in simulations. We can see in Fig.~\ref{fig:uncertainty}(b) that, for high-loss dielectrics, the measurement errors of $\epsilon_{ r}^{\prime\prime}$ are small. However, for low-loss dielectrics, the relative uncertainty significantly increases. This is because when the wave goes through a low-loss dielectric slab, the attenuation cannot be sufficiently accumulated, and the changes of $S$-parameters caused by material losses are not evident. In this case, uncertainties in $S$-parameters can easily cause inaccurate estimations  of the loss tangent. 
	This is the common shortcoming of the transmission/reflection method for measuring thin low-loss material samples \cite{catala2003accurate}. 
	For lossy dielectrics, the influences of material loss on the measured $S$-parameters are observable, and the imaginary part of permittivity can be estimated accurately.  
	
	Finally, we should note that the $S$-parameter uncertainties provided by the VNA manufacturer are their worst values.
	In reality, the perturbations of $S$-parameters are not so strong, and the uncertainties shown in Fig.~\ref{fig:uncertainty} might be overestimated. A good evidence is Fig.~\ref{fig:extraction_comparision_b} where the extracted imaginary part of permittivity does not fluctuate as strong as estimated.

	\section{conclusion}
	To summarize, this paper reports a fast and robust method to measure dielectric slabs in a rectangular waveguide junction. 
	The method does not require meticulous control of sample shape and position, which is particularly useful for millimeter-wave and sub-terahertz-wave measurements. 
	The physics behind this method is that, the equivalent shunt impedance of the waveguide junction is only related to the permittivity of measured material if the thickness of the sample is known in advance. 
	We develop an analytical formula to extract the permittivity of electrically thin materials ($d<\lambda_{ d}/10$). For thick dielectrics ($\lambda_{ d}/10<d<\lambda_{d}/2$), numerical tool is needed to extract the permittivity. The method can accurately retrieve the real part of permittivity, while the prediction of the imaginary part is accurate only for medium-loss and high-loss materials.

	\section{Acknowledgements}
	The authors would like to thank Francisco Cuesta for his help in glass sample measurements.
	
	\bibliography{references}
	\bibliographystyle{IEEEtran}

\end{document}